\documentclass{aastex}
\newcommand{\msun}{$M_\odot$} 
\newcommand{\Msun}{\mbox{$\,M_\odot$}}
\newcommand{\persec}{\mbox{$\second^{-1}$}}

\newcommand{\ppm}{\mbox{$\pm$}}

\newcommand{\cgslum}{\erg\persec}
\newcommand{\kmpersec}{\km\persec}
\newcommand{\counts}{\mbox{$\rm\ counts$}}
\newcommand{\countspersec}{\counts\persec}
\newcommand{\approxlt}{\mbox{$\lesssim$}}

\newcommand{\etal}{et~al.}
\newcommand{\eg}{e.g.}
\newcommand{\ie}{i.e.}
\newcommand{\cenx}{Cen~X-4}

\newcommand{\aql}{Aql~X-1}

\newcommand{\saxpsr}{SAX~J1808.4-3658}

\newcommand{\ee}[1]{\mbox{$10^{#1}$}}

\newcommand{\keV}{\mbox{$\rm\,keV$}}

\newcommand{\km}{\mbox{$\rm\ km$}}
\newcommand{\Hz}{\mbox{$\rm\ Hz$}}
\newcommand{\second}{\mbox{$\rm\ s$}}

\newcommand{\usec}{\mbox{$\rm\,\mu s$}}

\newcommand{\erg}{\mbox{$\rm\,erg$}}

\newcommand{\axaf}{{\em AXAF\/}}

\newcommand{\rxte}{{\em RXTE\/}}
\newcommand{\xmm}{{\em XMM\/}}

\newcommand{\citenp}{\citealt}
\newcommand{\citeyp}{\citet}
\slugcomment{\it ApJ, accepted}
\shorttitle{No Persistent Pulsations in \aql}

\begin{document}

\title{No Persistent Pulsations in \aql\ as it Fades into Quiescence}
\author{A. M. Chandler and R. E. Rutledge}
\affil{
Space Radiation Laboratory, California Institute of Technology, \\
MS 220-47, Pasadena, CA 91125}
\email{amc@srl.caltech.edu}

\begin{abstract}
We searched for coherent X-ray pulsations from \aql\ in 
a series of \rxte\ observations taken shortly after 
a recent outburst. During the course of these observations, 
\aql\ passes through an apparent ``propeller'' phase as its 
luminosity fades to its quiescent value. No pulsations were detected, 
and we place upper limits (ranging from $0.52\%$ to $9.0\%$) 
on the fractional RMS amplitude of any periodic signal 
contained in the various data sets searched. 
This result has implications for the geometry of the system, if 
the quiescent luminosity is due to continued low-level accretion. 
Alternatively, our result supports the idea that the 
quiescent luminosity may be due to thermal emission. 
\end{abstract}

\section{INTRODUCTION}

The origin of the quiescent emission observed from low-magnetic field
transient neutron stars (NSs) has posed something of a puzzle.  When
first discovered observationally (in the transient \cenx;
\citenp{jvp87}), the X-ray emission, fit with a blackbody spectrum,
was too faint to have originated over the entire surface of a 10~km
NS.  It was suggested that the emission was due to accretion over a
fraction ($\sim \rm{few}\%$) of the NS surface.  As additional quiescent NS
X-ray spectra were observed and fit with a blackbody spectrum
\citep{verbunt94,asai96a,asai96b}, the small implied emission areas 
seemed to confirm this view.

The cause of accretion anisotropy over the surface of a low B-field NS
has been purported to be the ``propeller effect''
\citep{ill75,stella86}, in which at low accretion rates, the magnetic
field increases in importance relative to the gravitational field in
determining the accretion geometry, perhaps expelling much of the
accretion from the system when the effect of the magnetic field is
comparable to or stronger than that of the gravitational field at the
radius where the Keplerian frequency is equal to the spin frequency of
the NS.  When this occurs, some fraction of the accretion may follow
the magnetic field lines to the magnetic poles which, if these are
offset from the rotational poles, could conceivably produce X-ray
pulsations.  The efficiency of such a propeller has been estimated
recently \citep{menou99}.  However, some observations of decreasing
torque with increasing luminosity may indicate deviations from this
standard magnetic accretion scenario in low mass X-ray binaries 
\citep{chakrabarty97}.  If, after an outburst, accretion continues
into quiescence then pulsations might be expected, particularly if the
NS can be shown to have gone through a ``propeller phase'', in which
the accretion geometry in the vicinity of the NS is dominated by the
magnetic field, rather than by the gravitational field. No pulsations
from these objects in quiescence have yet been reported, although they
have been searched for \citep{verbunt94,asai96b}.

Thus, until recently, the view of the quiescent luminosity of low-B
transient NSs has been as due to continued accretion over a fraction
of the surface of the NS, perhaps caused by modified accretion
geometry due to the effects of the magnetic field, which may therefore
give rise to pulsations in the X-ray intensity.  

A different view of the origin of the quiescent luminosity has
recently been described \citep{brown98}, in which nuclear reactions at
the base of the NS crust keep the NS core heated to temperatures
($\approx10^8\, \rm{K}$) sufficient to explain a large fraction, if not
all, of the quiescent luminosity of these objects (\ee{32-33}
\cgslum); in this manner, the quiescent luminosity can be produced in
the absence of active accretion.  The discrepancy in the emission area
can be explained as due to an incorrect blackbody assumption --
specifically, if accretion is shut-off, metals stratify in the NS
atmosphere \citep{alcock80,romani87}, producing a pure H photosphere,
in which the free-free opacity, which decreases with increasing
frequency, dominates, permitting higher E photons to originate at
greater depth, thus at higher temperatures
\citep{rajagopal96,zavlin96}.  This produces an emergent spectrum which
is spectrally harder than a simple blackbody, and -- when described as
a simple blackbody -- requires a higher $T$, and subsequently lower
emission area, to parameterize it.  Spectral fits with H atmosphere
spectra show that the observed X-ray spectra from the three such
objects for which data exists (\cenx, \aql, and 4U~1608-522) imply
emission areas consistent with a 10 km NS \citep{rutledge99}.  Thus,
both their quiescent luminosities and their emergent X-ray spectra can 
be explained in this manner, without invoking active accretion
during quiescence.  An attractive feature of this explanation is that
it accounts for the similar luminosities of these three objects,
without requiring their quiescent mass accretion rates (a product of
the propeller efficiency and mass accretion rate through the accretion
disk) to be serendipitously similar.  

It remains an open question if accretion (at a level of
$\dot{M}$\approxlt$10^{-12}$\msun\  yr$^{-1}$) occurs at all, contributing
some fraction of the quiescent X-ray luminosity.  There are few
observational means by which this may be investigated.  One such
possibility is to search for metal lines in the quiescent X-ray
spectrum, which, if present, would imply that metals are being fed
into the atmosphere at a rate greater than gravity can stratify it
(\citenp{bildsten92}, see \citenp{zavlin96} for X-ray spectra of
metallic atmospheres).  Such spectra can be investigated by the latest 
generation of X-ray spectroscopy missions (\axaf\ and \xmm). 
A second method is to search for
intensity variability during quiescence.  In particular, as described
above, if pulsational variability at a particular frequency were to be
found in quiescence, this would support the scenario of active
accretion along the magnetic field lines to the NS surface.
Alternatively, if the NS magnetic field were strong enough to significantly
affect the photospheric opacity ($>10^{10}$~G; cf. \citenp{zavlin95}),
this would produce apparent ``hot spots'' near the magnetic poles of
the NS which, if offset from the rotation axis, could give rise to
pulsations; however, it is thought that type-I X-ray bursting sources
have magnetic fields below this value. Clearly discovery of pulsations
or constraints on the pulsed fraction during X-ray quiescence
(following the onset of a ``propeller'') would provide useful limits
for models of quiescent NS emission.

The system \aql\ lends itself to such a search for pulsations.  During
the X-ray decline of a recent outburst, the intensity at first
decreased by one order of magnitude over 17 days, then abruptly
(beginning at $L_x\approx10^{36}\ \cgslum$) by 3 orders of magnitude
over 3 days, which was coupled with the sudden spectral hardening of
the X-ray intensity \citep{campana98,zhang98a}.  The X-ray intensity
stopped dropping at a luminosity $L_x\sim10^{33}\ \cgslum$,
comparable to the quiescent luminosity observed previously
\citep{verbunt94}. This was interpreted as evidence for a
propeller-phase, though this remains to be confirmed through
repeated observations of \aql\ in outburst. 
If indeed this behavior does indicate a propeller phase in \aql, it
offers the opportunity to search for pulsations during
the period when the NS magnetic field substantially alters
the accretion geometry. 

Here we have searched for, and not found, pulsations in \aql\ 
during and immediately following its supposed propeller phase. 
In Sec.~\ref{sec:analysis}, we describe our search method and 
the results of our analysis.  In
Sec.~\ref{sec:discuss} we discuss these results and our conclusions.

\section{ANALYSES}
\label{sec:analysis}

\subsection{Search Methodology}
\label{sec:method}

We used high time resolution data taken with \rxte/PCA 
\citep{pca}. 
The PCA detectors have a total geometric area of 6500 cm$^2$, and a
nominal energy range of 2-60 keV.  We selected for analysis four
observations, based on the results of \citeyp{zhang98b}.
These were listed by Zhang \etal\ as observations~8, 10, 11, plus a
later observation, 12, which was not analyzed by Zhang due to the
faintness of the source -- which made their spectral analysis
infeasible, but can still be useful in our search for
pulsations. Based on Zhang's results, the beginning of the
propeller-phase appears to be between observations~9 and 10, and so
our timing analyses of observations~10-12 are expected to reveal
pulsations created by the magnetic-field modified accretion geometry
(if any), while observation~8 is included simply for comparison
purposes while the source was moderately brighter.  We list details of
the observations in Table~\ref{tab:observations}.  Data were obtained from
the XTE archive at HEASARC.

For our data analysis, we used the standard FTOOLS/XTE v4.0.  Using
standard-2 formatted data, we extracted the X-ray spectrum from each
observation and subtracted background counts (estimated using
pcabackest tool), and examined the counts spectra.  Based on the
spectrum of the faintest observation (number 12), which becomes 
background dominated above $\sim12\keV$, we selected for our
analysis PHA bins 0-20, corresponding in energy roughly to 2-12 keV.
We then extracted individual photon events from data in E125 mode
(which has time resolution of $\sim125\mu s$), and corrected the
photon arrival times (time of arrival; TOAs) to the solar system
barycenter.

\citeyp{zhang98b} reported strong evidence for a $549\ \rm{Hz}$  
pulsation frequency, observed during $\sim10\ \rm{seconds}$ following 
a type-I X-ray burst from \aql. (It is possible that this may be 
the second harmonic of the NS spin frequency.) 
Similar pulsations have been observed from type-I bursts of 
5 other sources (1728-34, \citenp{strohmayer96}; 
1636-53, \citenp{zhang96}; 1731-260, \citenp{smith97}; 
a galactic center source, possibly MXB 1743-29, \citenp{strohmayer97}; 
and 1702-43, \citenp{markwardt99}). 
Rather than restrict our efforts to a small range of frequencies 
around the previously measured value, 
we chose to do a more general search for pulsations 
over the broad frequency range $0.5-1024\ \rm{Hz}$. 
This range includes all previously reported 
rotation frequencies for type-I X-ray bursters. 
We were able to do this more general search without a significant 
loss of sensitivity for reasons which we 
explain below, in section \ref{sec:sensitivity}. 

The \aql\ system consists of a NS and a companion star, both orbiting 
the system's center of mass with a period of about 19 hours 
(see section~\ref{sec:phasespace} below). As a result of this
orbital motion, the NS is accelerating along our line of sight. Any
pulsations will therefore be observed with a doppler shifted frequency
which is not constant in time. In a standard fast Fourier transform (FFT)
analysis, the spectral power resulting from such a signal may be
spread out over many frequency bins, drastically decreasing the
probability of detection. 
For example, consider a 550 Hz pulsar in a 19 hour circular 
orbit with a maximum projected velocity of 100\kmpersec. 
If we calculate a coherent power spectrum using $T=2048$ seconds of data, 
the power from this pulsar will be spread over as many 
as 70 independent frequency bins 
(for the worst case orbital phase, the observed pulsar frequency covers 
a range $\Delta f = 0.034\Hz$, giving $T\Delta f=70$~bins). 
To counteract this 
effect, we attempt to remove the acceleration of the pulsar signal in
the time domain before performing the FFT.  Because we do not have
complete knowledge of \aql's orbital parameters, we must cover the
possible orbital parameter space with a number of acceleration trials and
repeat the Fourier analysis for each. 

For our purposes, the orbit (assumed
circular) can be characterized by three independent parameters --- 
the orbital period $P_{\rm orb}$, 
the projected circular velocity $v$, and the orbital phase $x_0$ of
the NS at the start of the observation in question. 
The projected circular velocity is the
magnitude of the NS's circular velocity projected along our line of
sight, $v = v_{\rm NS}\sin{i}$ where $i$ is the angle
between the line of sight and the normal to the plane of the
orbit. Note that we could equally well have chosen the projected
orbital radius ($a\sin{i} = v P_{\rm orb}/2\pi$) in lieu of
this parameter. The orbital phase $x_0$ is
measured in cycles, and is therefore in the range $[0,1]$, with
$x_0=0$ corresponding to ${\rm longitude}=0$ (where the observed pulsar
frequency would be measured at its minimum).  The signal from a pulsar
in such an orbit would be observed with a doppler shifted frequency
\begin{equation}
\label{eq:doppler}
f(t)=f_0 \left[1-{{v}\over{c}}\cos{({{2\pi}\over{P_{\rm orb}}} t +
2\pi x_0)}\right]
\end{equation}
\noindent where $f_0$ is the rest frequency of the pulsar, $c$
is the speed of light, and $t$ is the elapsed time since phase $x_0$. 

Pulsar orbital acceleration searches are typically carried out
(see, for example, \citenp{anderson90}) by correcting data with an assumed constant acceleration
$f(t)=f_0+\dot f t$. This is applicable when the data cover only a
small part of the pulsar's orbit, or the pulsar is very strong. In the
first case, the short span of orbit is well approximated by a constant
acceleration, while in the second case, some spreading of the feature
in the power spectrum can be tolerated without the signal disappearing
into the noise.  For the \aql\ searches reported here, we assumed that
neither of these conditions was satisfied. Our acceleration searches
therefore fully correct for an assumed circular orbit $(P_{\rm orb}, v,
x_0)$.  As compared with the constant acceleration method, this
circular acceleration method may require more trial accelerations to
cover a given orbital phase space, but the detection significance is
greatly increased. This is because the circular method recovers
significantly more signal power in a single power spectrum bin, more
than making up for the increased number of trials (the detection
significance is exponential in recovered power, but only linear in the
number of trials).

After preparing an energy selected, barycentered photon TOA list, 
we proceeded with the
search method as follows.  First, we assumed particular values
for the orbital parameters (\ie, a trial acceleration) from within the
search phase space (of orbital period, velocity, and initial phase; see
Table~\ref{tab:phasespace}).  We corrected the TOAs for the assumed
acceleration by introducing a corrected time $\tilde t$, which is a
function of the original time $t$, such that the frequency as a
function of $\tilde t$ is constant. 
Equivalently, we require the integrated phase to be linear in $\tilde t$: 
\begin{equation}
x(t)-x_0=\int_0^t f(t') dt'=f_0 \tilde t.
\end{equation} 
Integrating equation~(\ref{eq:doppler}) for a given acceleration trial 
$(P_{\rm orb},v, x_0)$, we see that the $i^{th}$ TOA $t_i$ is corrected to: 
\begin{equation}
\label{eq:tildet}
\tilde t_i=t_i
+{{\beta}\over{\omega}}\left[\sin(2\pi x_0)-\sin(\omega t_i+2\pi x_0)\right]
\end{equation}
where $\beta = v/c$ and $\omega=2\pi/P_{\rm orb}$. 

These corrected TOAs were used to construct a time series which was then
FFTed and used to produce an estimate of the power 
density spectrum (PDS). The PDS
was searched for candidates (frequency bins containing statistically 
significant excess power). The process was
repeated for each acceleration trial.  The spacing of acceleration
trials in (orbital period, velocity, and phase)-parameter space was chosen
to allow sensitivity to as weak a signal as possible while keeping the
computational requirements reasonable.

\subsection{Determination of Searched Parameter Space}
\label{sec:phasespace}

Similar, but inconsistent, values for the orbital period of \aql\ 
have been published by two groups. 
Based on observations of \aql\ during outburst, \citeyp{chevalier91} 
measured the orbital period to be $18.97\ppm0.02\ {\rm hr}$. 
\citeyp{shahbaz98} determined the orbital period in quiescence to be 
$19.30\ppm0.05\ {\rm hr}$. In a more recent analysis, \citeyp{chevalier98} 
report the period to be $18.9479\ppm0.0002\ {\rm hr}$, again measured 
during outburst, but they also report a quiescent period within $0.02\%$ of 
this value. 
To be safe, we chose to cover a range of orbital periods that 
encompasses a $3\sigma$ range in all of these measurements (as well 
as the periods in between): 
\begin{equation}
\label{eq:Pspace}
18.91 < P_{\rm orb} < 19.45\ {\rm hr}. 
\end{equation}
Fortunately, we were able to cover this entire range with one trial 
value of $P_{\rm orb}$ (see section~\ref{sec:searches} below). 

The projected circular velocity $v$ has not been measured. 
We can determine an upper limit on $v$ for our search by assuming 
a value for the NS mass, $m_{\rm NS}=1.4\Msun$ 
($\Msun =\rm{one\ solar\ mass}$), and choosing a 
maximum companion mass to which we will be sensitive. We can then 
use our knowledge of the orbital period to calculate the orbital 
velocity of the NS. 
Recently, the true optical counterpart of \aql\ has been identified 
as a late K type star \citep{chevalier99}. 
Again, to be safe we decided to cover a range of velocities corresponding 
to a companion star mass as high as $1.0\Msun$ (for all possible orbital 
inclination angles). This results in a search range of  
\begin{equation}
\label{eq:vspace}
0 \le v \le 130\kmpersec. 
\end{equation}
Note that our search was also sensitive to 
larger companion masses in a restricted range of inclination angles; \eg\ 
our search covered companion masses up to $2.0\Msun$ for 
$0^\circ \le i \le 40^\circ$. 

For the starting orbital phase for each observation, rather than rely 
on a particular ephemeris, we simply chose to search the full range 
\begin{equation}
\label{eq:xspace}
0 \le x_0 \le 1\ \rm{cycle}. 
\end{equation}
Our search phase space is summarized in Table~\ref{tab:phasespace}.

\subsection{Searches Performed}
\label{sec:searches}

The \rxte\ observations included in this analysis were roughly 9-13
kiloseconds in duration. The TOA data in each observation are broken
up into two or three cohesive sections, separated by gaps due to earth
occultations. Each continuous section is at least 2048 seconds
long, and we chose to study these continuous blocks individually. 
Observation~10, with one occultation drop-out, is therefore divided into 
two sections, which we call 
observations~10a and 10b. Observations~11 and 12 are similarly divided 
(into 11a, 11b, 11c, 12a, and 12b). We analyzed only the first section 
of observation~8, for a total of eight separate data sets. 
The photons from a given data set 
were binned into a $2^{23}$~point time series, with a time 
resolution of $\Delta t=244\usec$. 
Over the course of the observations, the source count rate
decreases from about 670\countspersec\ at the beginning of observation~8
to less than 10\countspersec\ by the end of observation~12, with a
background rate of about 31-37\countspersec\ (see Table~\ref{tab:results}). 

Our search method was described above in section~\ref{sec:method}. 
We now describe in detail the method used to determine the grid of trial 
values for the orbital parameters. 
Since we must discretize a continuous phase space, there will likely be 
some offset between a signal's actual parameters and our nearest trial values. 
The effect of this offset will be some spreading out of the signal 
power in the power spectrum, since the time dependence of the signal's 
frequency will not have been completely removed. In choosing the orbital 
trials, the general idea was to space them just finely enough 
to keep the remaining frequency drift within tolerable limits (to be 
quantified below). 

We began analytically, and then made some empirical adjustments. 
Considering each of the three orbital parameters ($P_{\rm orb},v,x_0$) 
separately, we took partial derivatives of equation~(\ref{eq:doppler}), 
to determine the frequency drift that results from offsets in the 
individual parameters. For example, an error in the velocity 
($\partial v=c\partial \beta$) causes a drift of 
\begin{equation}
\label{eq:partial_v}
\partial f = f_0 \cos{(\omega t+2\pi x_0)} \;\partial \beta.
\end{equation}
We are interested only in the extrema 
of the frequency range covered by the signal over the course of the 
observation time $T$, and for simplicity, we chose to make our velocity 
trial values independent of the other parameters. We therefore replace 
the cosine factor in equation~(\ref{eq:partial_v}) by its maximum change 
over the course of an observation ($T=2048\ \rm{s}$), and we replace 
$f_0$ by our maximum search frequency 
($f_0\rightarrow1024\ \rm{Hz}$). The maximum frequency drift caused by 
a velocity offset is therefore 
\begin{equation}
\label{eq:maxdrift_v}
\partial f_{\rm max}=1024\ 0.19\ \partial\beta\ \rm{(Hz)}. 
\end{equation}
A ``tolerable'' drift must be no greater than the independent 
frequency resolution of the power spectrum, $F$. Since we are considering each 
orbital parameter separately, for these initial calculations we restrict 
the error-induced drift from each to $F/2$. This is somewhat 
arbitrary, and the actual spacings used were decided on with the 
help of simulations. Note that the allowable 
spacing of the trials can be twice the maximum allowable offset. 
We now have for the velocity trial spacing
\begin{equation}
\label{eq:spacing_v}
\Delta v_{\rm trial}\lesssim{{cF}\over{1024\ 0.19}}
=0.77\;\left({{F}\over{5\times10^{-4}\ \rm{Hz}}}\right)\kmpersec. 
\end{equation}
For a coherent FFT of 2048 seconds of data, the independent Fourier 
resolution is $F=1/T=4.88\times10^{-4}\ \rm{Hz}$. 
The search strategy that we settled on actually used a larger 
frequency resolution (see below); for now we will 
leave the expressions for the trial spacings in terms of $F$ explicitly. 

Again, for simplicity, we chose to keep the spacing of the $P_{\rm orb}$ and 
$x_0$ trials independent of $P_{\rm orb}$ and $x_0$, but we did allow for 
velocity dependence. For the orbital period, we ultimately find 
\begin{equation}
\label{eq:spacing_p}
\Delta P_{\rm trial}\lesssim2800\;
\left({{F}\over{5\times10^{-4}\ \rm{Hz}}}\right)\;
\left({{v}\over{100\kmpersec}}\right)^{-1}\ \rm{seconds}.
\end{equation}
Even for the smallest 
Fourier resolution ($F=4.88\times10^{-4}\ \rm{Hz}$) and the largest 
velocity in our search range ($v=130\kmpersec$), the orbital 
period trial spacing turns out to be larger than our target 
search range of 
$(19.45\ \rm{hr}-18.91\ \rm{hr})\ 3600\ \rm{s\ hr^{-1}}=1944\ \rm{s}$. 
In other words, a single trial value was sufficient to cover 
our entire search range in $P_{\rm orb}$. 
Thus, {\em we did not actually search over 
trial values of the orbital period.} 
For the initial orbital phase trial spacing, we find 
\begin{equation}
\label{eq:spacing_x}
\Delta x_{\rm trial}\lesssim1.2\times10^{-3}\;
\left({{F}\over{5\times10^{-4}\ \rm{Hz}}}\right)\;
\left({{v}\over{100\kmpersec}}\right)^{-1}
\ \rm{cycles}.
\end{equation}

If we were to use these spacings with coherent 2048 second FFTs, 
each of the eight data sets would require over 95\,000 trial accelerations. 
This would require $\gtrsim100$ days of CPU time on a 
Sun Ultra 10 workstation. 
To reduce the computational requirements, we chose to utilize incoherently 
stacked power spectra. The original data of duration $T$ are divided 
into ${\cal S}$ sections of duration $T/{\cal S}$. 
Each section is FFTed individually 
and the ${\cal S}$ individual power spectra are (incoherently) added together. 
The independent Fourier step size has been increased by a factor of 
${\cal S}$ ($F=1/(T/{\cal S})={\cal S}/T$) and the trial parameter spacings 
increase by the same factor. The total number of trials is 
therefore reduced by a factor of ${\cal S}^2$ (since we are searching over 
two parameters). The reduction in the number of trials comes at the cost of 
reduced sensitivity, so the number of stacks ${\cal S}$ should be kept 
as small as possible, just barely bringing the number of trials 
to a feasible level. For our analyses, we chose to use ${\cal S}=4$. 

With the above analytical calculations as guides, we used simulations 
to determine the trial spacings actually used in the search. 
Our simulations upheld the decision to use a single $P_{\rm orb}$ trial. 
Since there are indications that the orbital period is more likely 
to lie towards the lower end of our search range ($18.91-19.45$ hr), 
we chose to use $P_{\rm orb}=19.00\ \rm{hr}$ in our trials. We decided to 
use a velocity trial spacing of 
\begin{equation}
\Delta v_{\rm trial}=2.0\kmpersec.
\end{equation}
This is about $1.5\times$ finer than the spacing calculated above 
(recall that we are using ${\cal S}=4$ stacks). For the orbital 
phase, we decided on a trial spacing of 
\begin{equation}
\Delta x_{\rm trial}=0.011\;\left({{v}\over{100\kmpersec}}\right)^{-1}
\ \rm{cycles},
\end{equation} 
which is about $2.3\times$ coarser than the spacing indicated 
in the analytical calculation. For small velocities, we used a minimum 
of 16 $x_0$ trials (except for the single $v=0$ trial). 

Using these spacings with 
$2\times$ oversampled power spectra (exactly like our actual searches), 
we determined that at least $98\%$ of the 
time, we were able to recover at least $77\%$ of a simulated signal's 
power in a single spectral bin, even in the least favorable 
regions of our search phase space. 
We simulated signals whose orbital parameters were offset from our 
search trials and whose pulsation frequencies were offset from our 
discrete Fourier frequencies. 
The reduction in recovered signal power is due to a combination of these 
factors. 
(Note that it is mere coincidence that a simple FFT recovers, on average, 
$77\%$ of a signal's power based solely on offset from the discrete 
Fourier frequencies. In a $2\times$ oversampled spectrum, the {\em minimum} 
power recovered, based solely on frequency offset, is $81\%$.) 
Our peak detected power was not always in the frequency 
bin closest to the rest frequency of the simulated pulsar. This has no effect 
on the detection of pulsations; the true frequency and orbit of 
the pulsar can be refined after the initial detection. 

In total, we covered the orbital phase space with 
4069 acceleration trials. The same accelerations were used for each of 
the eight data sets.

\subsection{Estimation of Detection Sensitivities}
\label{sec:sensitivity}

To characterize the sensitivity of our search, we wish to place 
quantitative limits on the minimum signal strength required for 
a significant detection. Since the noise statistics of our 
power spectra are well understood, it is a simple matter to 
determine the detection threshold, the minimum spectral power 
$P_{\rm det}$ required for a detection. 
This information can be used to determine the 
minimum required signal strength. For a review of detection thresholds, 
detection sensitivities, and upper limits, see \citet{vaughan94}. 

Our stacked power spectra were constructed by summing four individual 
spectra, each normalized to a mean power of one (note that this 
convention differs from the commonly used Leahy normalization 
\citep{leahy83}, for which the mean noise power is two). 
In the absence of a signal, the power $P$ in a given spectral bin 
follows a $\chi^2$~distribution. Specifically, $2P$ is 
$\chi^2$~distributed with $2{\cal S}=8$ degrees of freedom. 
The probability that the power $P$ in a single PDS bin will exceed a 
given value $P_0$ is therefore 
\begin{equation}
p(P>P_0)=e^{-P_0} \left(1+P_0+{{1}\over{2}}P_0^2+{{1}\over{6}}P_0^3\right).
\end{equation}

Considering each frequency bin of each PDS to be one 
search ``trial,'' our entire search consisted of 
$N_{\rm trials}=3.4\times 10^{10}$ such trials. 
This is the product of the number of observations searched (8), 
the number of acceleration trials for each observation ($4069$), 
and the number of frequencies searched in each power spectrum ($2^{20}$). 
The latter number is equal to the frequency range searched ($1024\Hz$) 
divided by the independent frequency resolution ($F=({\cal S}/2048)\Hz$) 
times the oversampling factor ($2$). 
Of course, neighboring bins in an oversampled 
spectrum are not independent, and neighboring acceleration trials 
may not produce truly independent spectra. Thus $N_{\rm trials}$ 
is an upper limit to the number of statistically independent trials 
in the search. Since we are overestimating 
the number of independent trials, our detection threshold will 
be conservative. 

The statistical significance $S$ of a measured power $P_0$ is equal to the 
probability that the power was produced by a random 
noise fluctuation. For small $p$ (large powers), the significance is 
\begin{equation}
S=N_{\rm trials}\  p(P>P_0).
\end{equation} 
To achieve our target significance of $10^{-4}$, we require a PDS power 
of at least 
\begin{equation}
P_{\rm det}= 43.
\end{equation}

For a given data set containing $N_t$ total counts, $N_s$ of which 
were emitted by the source, 
we can relate the source strength to the expected spectral power by 
\begin{equation}
\label{eq:expectP}
\langle P\rangle = 4+{{1}\over{4}} {{N_s^2}\over{N_t}} {\cal F}^2 
\end{equation}
\citep{buccheri87,mvdk89} 
where ${\cal F}$ is the pulsed fraction, \ie\ the fraction 
of source counts that actually contribute to the pulsation. 
The first term in equation~(\ref{eq:expectP}) is the expected noise 
power ($\langle P_N\rangle=4$ because we have summed four power spectra 
each normalized to unity), while the second term is the expected 
signal power. Here, we have assumed that the signal waveform is 
sinusoidal. 
$N_s$ and $N_t$ are determined for each data set before beginning 
the pulsation search. Therefore, the sensitivity of 
each search is determined as a limit on the pulsed fraction ${\cal F}$. 

We now calculate the detection sensitivities for the eight 
data sets searched. For a given observation, the detection sensitivity 
is expressed as the minimum pulsed fraction required to produce 
a spectral power exceeding $P_{\rm det}$ with high confidence. 
Instead of simply solving equation~(\ref{eq:expectP}) for ${\cal F}$ 
using the detection threshold power $P_{\rm det}=43$, 
we must allow for statistical variation of the spectral power 
produced by a source of a given strength, as well as variation in the 
recovered power due to the discrete nature of the search trials. 

To determine the detection sensitivity, we must consider the probability 
distribution of power in a spectral bin containing a signal plus noise. 
If, in the absence of noise, the signal produces a power $P_{\rm sig}$, 
then the power $P$ in the presence of noise will be distributed according to 
\begin{equation}
\label{eq:groth}
p_n(P>P_0;P_{\rm sig})=\exp[-(P_0+P_{\rm sig})] \sum_{m=0}^{\infty} 
\sum_{k=0}^{m+n-1} P_0^k P_{\rm sig}^m/(k!m!)
\end{equation}
\citep{groth75}, where $n$ is the number of independent bins that were 
summed to produce $P$ (\ie, in our case $n={\cal S}=4$). 
Thus, $p_n(P>P_0;P_{\rm sig})$ is the probability that the power 
will exceed $P_0$ in the spectral bin containing the signal. 
It is important to note that noise power and signal power are not simply 
additive; $P$ will not always exceed $P_{\rm sig}$ (see \citet{vaughan94} 
for a discussion of this point). For a given $N_s$ and $N_t$, we would like 
to find the pulsed fraction ${\cal F}$ that is $95\%$ likely to 
produce power $P>P_{\rm det}$. 

To ensure that our reported sensitivity is as conservative as possible, 
we will not naively invert equation~\ref{eq:groth} for $P_{\rm sig}$ 
with $P_0=P_{\rm det}$. Instead we consider the worst case scenario 
covered by our search parameter space. Due to the discrete binning 
of photons in the construction of the time series, the recovered signal 
power in the FFT falls off with increasing frequency. A signal near our 
maximum search frequency --- $1024\Hz$, which is half the Nyquist 
frequency of our power spectra --- will produce $81\%$ of the power 
that would be recovered from a low frequency signal of the same 
intrinsic strength. Also, due to our discrete grid of acceleration trials 
and frequency trials, we are only $98\%$ likely to recover more than 
$77\%$ of a signal's available power. Thus, in our worst case, we 
are $98\%$ likely to recover at least $(0.77)(0.81)P_{\rm sig}= 
0.62 P_{\rm sig}$. 

We account for this reduction in recovered signal power 
approximately by solving equation~\ref{eq:groth} in the form 
\begin{equation}
\label{eq:groth2}
p_n(P>P_{\rm det}; 0.62 P_{\rm sig})=0.97. 
\end{equation}
The probability of our recovered signal power exceeding $0.62 P_{\rm sig}$ 
{\em and} our total spectral power $P$ exceeding $P_{\rm det}$ is then 
approximately $(0.98) (0.97) \approx 0.95$, 
giving our desired $95\%$ confidence. 

Solving equation~\ref{eq:groth2} (numerically) results in 
$P_{\rm sig}=93.6$. For a given observation, the minimum required 
pulsed fraction is then given by ${\cal F}=(4 N_t P_{\rm sig}/N_s^2)^{1/2}$ 
and converted to RMS by dividing by $2\sqrt{2}$. Thus, our 
$95\%$ confidence detection sensitivities 
for observations 8a, 10a, 10b, 11a, 11b, 11c, 12a, and 12b 
are given by $0.60\%$, $1.2\%$, $1.2\%$, $2.7\%$, $2.9\%$, $3.0\%$, 
$5.6\%$, and $10\%$ respectively. We verified these numbers with Monte Carlo 
simulations. 

Our search included spin frequencies up to $1024\Hz$. 
Had we restricted our search to a smaller range of frequencies, 
corresponding to the previously reported pulsations, our detection 
sensitivities would not have been drastically different. To achieve the same 
detection significance, with far fewer trials, we require only 
a slightly reduced spectral power. And 
since the pulsed fraction limit depends essentially on the 
square root of the required power, 
the detection sensitivities are not terribly sensitive to the number 
of frequency 
trials in the search. Given Fox's best estimate for the pulsation 
frequency $549.76^{+0.05}_{-0.03}\Hz$ (D. Fox 1999, private communication), 
if we had confined 
our search to a $5\sigma$ range (about $549.76\Hz$ and $274.88\Hz$), 
our detection sensitivity in observation~10a 
would have improved to $1.0\%$, while in observation~12b the limit would have 
been $8.8\%$.

\subsection{Verification of Procedures}
\label{sec:verification}

Simulated data were used to test our search codes and to verify the
sensitivity of the search.  As an additional check, we applied our
acceleration method to \saxpsr\ --- the accreting millisecond pulsar
\citep{rudi98,deepto98}.  Since our search was for a similar pulsar, with
$\sim$~few hundred Hz frequency in a $\sim$~few hour orbit about a low-mass
companion, \saxpsr\ 
provides a useful test of our acceleration method.

We analyzed an \rxte/PCA observation (observation
\#30411-01-01-04).  We selected energy PHA bins corresponding to
0.4-17 keV from data of the same type we use in our search (E125).  We
barycenter corrected the arrival times with the position given by 
\citet{roche98}, and used 2048 seconds of data beginning at 
Mission Elapsed Time (MET, seconds since 1994.0) 
135395082. 

The pulsations in these data are easily detected --- the $\sim401$~Hz
signal is obvious in an unaccelerated PDS, with significant power
detected in each of about 30 adjacent frequency bins, using 2048
seconds of data (Figure~\ref{fig:1808}, dotted line).  The highest
single bin power represents about $22\%$ of the total signal power in
the extended feature.

Using the known orbital parameters \citep{deepto98} to remove the
acceleration from the data, the spreading of the signal power is 
reduced to about 6 independent Fourier bins 
around the correct rest frequency of the pulsar,
400.975\,210\,6(8)~Hz, with $\sim85\%$ of the total power in a single
spectral bin (Figure~\ref{fig:1808}, solid line).  Oversampling reveals that 
the spreading that remains is
consistent with the expected sinc-function response of the (discrete) 
FFT to a signal with the pulsar's rest frequency \citep{mvdk89}.  Thus,
we confirm that our technique successfully removes the effect of the
doppler shift due to the pulsar's orbital motion.

\subsection{Search Results}
\label{sec:results}

No candidate pulsations from any of the \aql\ observations exceeded our
predetermined detection threshold. Using our largest detected search 
power, we can calculate upper limits 
on the strength of any pulsar signal contained in the
various data sets searched \citep{vaughan94}, provided its orbit 
and spin frequency are covered by our search phase space. 
These upper limits are more restrictive than the {\em a priori} 
detection sensitivities. 

The upper limit calculation is essentially the same as the 
detection sensitivity calculation, with the detection threshold power 
$P_{\rm det}$ replaced by the maximum observed power $P_{\rm max}=30.96$. 
The results are shown in Table~\ref{tab:results}. 
For example, with $95\%$ confidence, we can say that there is no 
sinusoidal pulsar signal in observation~10a with a fractional RMS 
amplitude greater than $1.0\%$. 
The limits we can place on the fractional RMS 
become less stringent as the source flux decreases, increasing to an 
upper limit of $9.0\%$ during observation~12b, when \aql\ was observed to
be faintest ($\sim$1~mCrab; $13\,000$ PCA\countspersec=1 Crab).

\section{DISCUSSION}
\label{sec:discuss}

We find no evidence of pulsations from \aql\ as it fades into
quiescence. The upper limits on the pulsed fraction for the eight 
data sets searched range from $0.52\%$ to $9.0\%$ RMS. 
If, indeed, during the period identified as the
``propeller phase'' \citep{campana98,zhang98a} the NS magnetic field
significantly modifies the accretion flow geometry in the vicinity of
the NS, the apparent absence of pulsations does not support the
hypothesis that the quiescent emission is due to continued accretion. Our
results cannot completely rule out such accretion, however, since it
is possible that the geometry of the system may not lead to detectable
pulsations; for example, the magnetic axis may be very nearly aligned
with the rotation axis, or the rotation axis may point directly
towards the earth.  The non-detection of pulsations coupled with the
observation of a thermal spectrum during quiescence \citep{rutledge99}
favors the interpretation that the quiescent luminosity is not due to
accretion, but rather to a hot NS core \citep{brown98}.

\acknowledgements
We thank Stuart B. Anderson and Thomas A. Prince for helpful 
discussions. We also thank Lars Bildsten, Sergio Campana, and an 
anonymous referee for valuable comments on the 
manuscript. This work was supported by NASA Grant \#NAG5-3239.

\clearpage
\bibliographystyle{apj}

\clearpage

\begin{figure}[htb]
\plotone{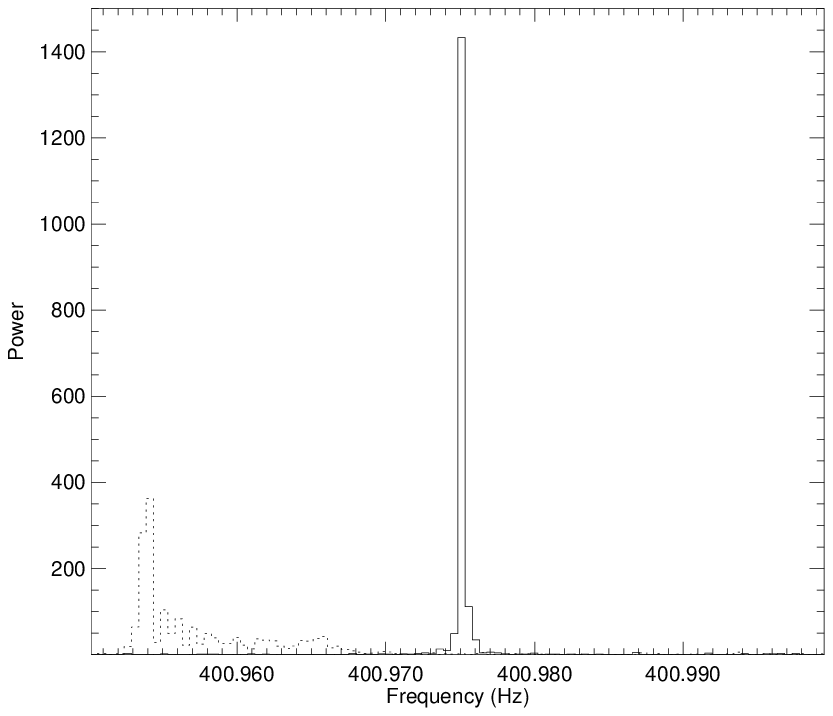}
\end{figure}

\figcaption[f1.eps]{
\label{fig:1808}
\saxpsr. The dotted line shows a section of the unaccelerated power spectrum. 
The spreading of the signal power due to the orbital doppler shift is evident. 
The solid line shows the 
power spectrum after removal of the doppler shift using our 
acceleration technique. The signal power has not been confined to a 
single bin because the pulsar frequency lies between two discrete Fourier 
frequencies. During this observation, \saxpsr\ passes through
${\rm longitude}=0$. 
The peak in the uncorrected spectrum is therefore maximally offset 
from the pulsar's rest frequency, 
$\Delta f_{\rm max}=\beta f_0=0.022\Hz$. 
}

\clearpage

\begin{table}[htb]
\begin{center}
\caption{\rxte\ \aql\ Analyzed Observations \label{tab:observations}}
\begin{tabular}{ccrr} \tableline \tableline
Observation	& Start Time / End Time			 & \multicolumn{1}{c}{Start Time} 	& \multicolumn{1}{c}{Start Time}   \\ 
		& (UT)					 & \multicolumn{1}{c}{(JD)}		&\multicolumn{1}{c}{ (MET) }       \\ \tableline
20098-03-08-00  & 1997-03-01 21:33 1997-03-02 00:21	 & 2450509.398				& 99869572     \\
20098-03-10-00  & 1997-03-05 22:55 1997-03-06 01:28    	 & 2450513.456				& 100220227    \\
20098-03-11-00  & 1997-03-08 21:31 1997-03-09 01:11   	 & 2450516.394				& 100474022    \\
20098-03-12-00  & 1997-03-10 21:25 1997-03-10 23:55   	 & 2450518.389				& 100646431    \\ \tableline
\end{tabular}
\end{center}
\end{table}

\clearpage

\begin{table}[htb]
\begin{center}
\caption{Searched Parameter Space\label{tab:phasespace}}
\begin{tabular}{lr} \tableline \tableline
Parameter & Parameter Space \\ \tableline
$P_{\rm orb}$ 		& $18.91-19.45\ \rm{hr}$		\\
$v_{\rm NS}\sin{i}$	& $0-130\kmpersec$		\\
Initial Orbital Phase 	& $0-1\ \rm{cycles}$	\\
Pulsar Spin Frequency	& $0.5-1024\ \rm{Hz}$		\\ \tableline
\end{tabular}
\end{center}
\end{table}

\clearpage

\begin{table}[htb]
\begin{center}
\caption{Observation Parameters and Search Results \label{tab:results}}
\begin{tabular}{ccccc} \tableline \tableline
		& MET      & Source	& Background	& Fractional RMS \\ 
Observation	& Analyzed & Count Rate\tablenotemark{a}& Count Rate\tablenotemark{a,b}& Amplitude\tablenotemark{c}	\\ \tableline
8a		& 99869572-98871620   & 670.6\ppm0.8	& 37.3		& $<0.52\%$		\\
10a		& 100220227-100222275 & 188.2\ppm0.6	& 35.8		& $<1.0\%$		\\
10b		& 100226227-100228275 & 177.4\ppm0.6	& 35.8		& $<1.1\%$		\\
11a		& 100474022-100476070 & 52.6\ppm0.5	& 35.7		& $<2.3\%$		\\
11b		& 100479522-100481570 & 48.6\ppm0.5	& 35.7		& $<2.5\%$		\\
11c		& 100485082-100487130 & 45.5\ppm0.5	& 35.7		& $<2.6\%$		\\
12a		& 100646431-100648479 & 19.0\ppm0.5	& 31.2		& $<4.9\%$		\\
12b		& 100652431-100654479 & 9.2\ppm0.5	& 31.2		& $<9.0\%$		\\ \tableline
\end{tabular}
\tablenotetext{a}{counts s$^{-1}$ ($13\,000\countspersec = 1\ \rm{Crab}$)}
\tablenotetext{b}{Systematic uncertainty $\sim0.5\countspersec$}
\tablenotetext{c}{Upper limits are 95\% confidence}
\end{center}
\end{table}

\end{document}